\documentclass[]{article}
\usepackage{packages}

\usepackage{arxiv}

\title{Agent-Based Analysis of the Impact of Near Real-Time Data and Smart Balancing on the Frequency Stability of Power Systems}

\date{March 26, 2025}	

\author{ 
	\href{https://orcid.org/0009-0009-1660-9631}{\includegraphics[scale=0.06]{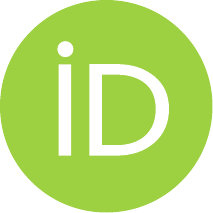}\hspace{1mm}Johannes Lips} \\
	Institute of Combustion and Power Plant Technology\\
	University of Stuttgart\\
	Stuttgart, Germany \\
	\texttt{johannes.lips@ifk.uni-stuttgart.de} \\
        \And
        \href{https://orcid.org/0009-0009-9839-5103}{\includegraphics[scale=0.06]{orcid.pdf}\hspace{1mm}Boyana Georgieva} \\
	Institute of Combustion and Power Plant Technology\\
	University of Stuttgart\\
	Stuttgart, Germany \\
	\texttt{boyana.georgieva@ifk.uni-stuttgart.de} \\
	\And
        Dominik Schlipf \\
	\phantom{Institute for Com} TransnetBW GmbH \phantom{Institute for Com}\\
	Stuttgart, Germany \\
	\And
	\href{https://orcid.org/0000-0002-0208-4100}{\includegraphics[scale=0.06]{orcid.pdf}\hspace{1mm}Hendrik Lens} \\
	Institute of Combustion and Power Plant Technology\\
	University of Stuttgart\\
	Stuttgart, Germany \\
}

\renewcommand{\shorttitle}{Agent-Based Analysis of the Impact of Near Real-Time Data and\\ Smart Balancing on the Frequency Stability of Power Systems}

\hypersetup{
pdftitle={Agent-Based Analysis of the Impact of Near Real-Time Data and Smart Balancing on the Frequency Stability of Power Systems},
pdfauthor={J. Lips, B. Georgieva, D. Schlipf, H. Lens},
pdfkeywords={smart balancing, power system stability, energy market design, frequency restoration reserve, imbalance settlement,  passive balancing },
}

\begin{document}
\pgfkeys{/pgf/number format/.cd,1000 sep={\,}}
\maketitle

\begin{abstract}
Single imbalance pricing provides an incentive to balance responsible parties (BRPs) to intentionally introduce power schedule deviations in order to reduce the control area imbalance and receive a remuneration through the imbalance settlement mechanism.
This is called smart balancing or passive balancing and is actively encouraged in, e.g., the Netherlands and Belgium through the publication of near real-time (NRT) data on the control area imbalance by the transmission system operator.
It is known that under certain conditions, smart balancing can deteriorate the frequency stability of the power system.

This paper examines how the publication of different types of NRT data affects smart balancing and the frequency stability.
A Monte-Carlo simulation of a dynamic multi-agent model is performed to analyse the effects of smart balancing with different parameters for the agents and the environment, using historical time series of the power imbalance of the German control block as a basis.

It is found that smart balancing can significantly reduce the amount and cost of frequency restoration reserve activation, but leads to a general increase of the frequency variability.
Depending on the type of NRT data and agent parameters, the frequency stability margins are also reduced.
The negative effects on the frequency stability are stronger when NRT data is published using large bins and with long delays.
\end{abstract}


\section{Introduction}\label{sec:intro}
\subsection{Imbalance Settlement and Smart Balancing}\label{sec:intro_smart}

In Europe, each \gls{ca} of a synchronous area is scheduled to be balanced for each 15-minute \gls{isp}.
This means that scheduled energy generation and imports equal scheduled consumption and exports in each \gls{isp} \citep{entsoeBalGL2018}.
The energy balance between supply and demand is achieved through electricity markets, in which \glspl{brp} trade energy, resulting in a schedule for each \gls{brp}.
\Glspl{brp} should follow their schedule as best they can, but typically have schedule deviations due to inevitable disturbances or forecast errors of the renewable generation or the load.
This means that they produce more or less energy than scheduled for that \gls{isp}.
The sum of the imbalances of all \glspl{brp} $b$ in a \gls{ca} is the \gls{ca} imbalance, 
\begin{equation}
    E_\mathrm{ACE} = \sum_{b\in\mathrm{CA}} E_b \,,
\end{equation}
with \glsunset{ace}\gls{ace} standing for \glsentrylong{ace}.
The \gls{ace} needs to be compensated for by the activation of balancing energy in the form of \glspl{frr}.

\begin{table*}[!t]   
\begin{framed}
\nomenclature[A]{TSO}{transmission system operator}
\nomenclature[A]{CA}{control area}
\nomenclature[A]{BRP}{balance responsible party}
\nomenclature[A]{ACE}{area control error}
\nomenclature[A]{ISP}{imbalance settlement period}
\nomenclature[A]{(a)FRR}{(automatic) frequency restoration reserve}
\nomenclature[A]{FCR}{frequency containment reserve}
\nomenclature[A]{NRT}{near real-time}
\nomenclature[A]{GCC}{German grid control cooperation}
\nomenclature[A]{RMSE}{root-mean-square error}

\nomenclature[B]{\E}{Exact NRT data}
\nomenclature[B]{\Es}{Exact NRT except for $[-120,120]\,\si{MW}$}
\nomenclature[B]{\El}{Exact NRT except for $[-900,970]\,\si{MW}$}
\nomenclature[B]{\Is}{Interval NRT data}
\nomenclature[B]{\Il}{Interval NRT except for $[-900,970]\,\si{MW}$}

\nomenclature[S]{$E$}{imbalance energy [MWh]}
\nomenclature[S]{$E_\mathrm{FRR}$}{FRR balancing energy [MWh]}
\nomenclature[S]{$b$}{index used for BRPs [-]}
\nomenclature[S]{$(\Delta)f(t)$}{grid frequency (deviation) [Hz]}
\nomenclature[S]{$T_\mathrm{NRT}$}{NRT discretization time step [s]}
\nomenclature[S]{$T_\mathrm{sim}$}{end time of the simulation [s]}
\nomenclature[S]{$N_\mathrm{NRT}$}{number of elements in vector signals [-]}
\nomenclature[S]{$k$}{discrete time variable [-]}
\nomenclature[S]{$t$}{continuous time variable [s]}
\nomenclature[S]{$P_\mathrm{demand}(t), \bm{P}_\mathrm{demand}$}{FRR demand [MW]}
\nomenclature[S]{$\bm{l}_k$}{lookahead/`perfect forecast' signal [MW]}
\nomenclature[S]{$\bm{d}_k$}{NRT imbalance data [MW]}
\nomenclature[S]{$\overline{\bm{d}}_k,\underline{\bm{d}}_k$}{upper and lower interval of NRT imbalance data [MW]}
\nomenclature[S]{$P_\mathrm{d}(t), \bm{P}_\mathrm{d}$}{disturbance[MW]}
\nomenclature[S]{$\hat{\bm{x}}$}{$\bm{P}_\mathrm{demand}$ estimation [MW]}
\nomenclature[S]{$\overline{\hat{\bm{x}}},\underline{\hat{\bm{x}}}$}{robustness interval for $\hat{\bm{x}}$ [MW]}
\nomenclature[S]{$u_b(t), \bm{u}_{b,k}$}{smart balancing decision of BRP $b$ at times $t$ and $k$ [-]}
\nomenclature[S]{$y_b(t),\bm{y}_{b,k}$}{activated smart balancing power of BRP $b$ at times $t$ and $k$ [MW]}
\nomenclature[S]{$P_\mathrm{smart}(t)$}{total activated smart balancing power [MW]}
\nomenclature[S]{$T_\mathrm{delay}$}{delay in the NRT data publication [s]}
\nomenclature[S]{$\bm{\mu}$}{mean of a distribution [MW]}
\nomenclature[S]{$\bm{\Sigma}$}{covariance matrix [MW$^2$]}
\nomenclature[S]{$\hat{\bm{\sigma}}$}{estimated standard deviation [MW]}
\nomenclature[S]{$\mathcal{TN}$}{truncated multivariate normal distribution}
\nomenclature[S]{$\mathcal{N}$}{multivariate normal distribution}
\nomenclature[S]{$c$}{imbalance price function [EUR/MWh]}
\nomenclature[S]{$C$}{smart balancing profit function [EUR]}
\nomenclature[S]{$c_\mathrm{smart}$}{activation cost function [EUR]}
\nomenclature[S]{$\bm{\delta}_j$}{smart balancing decision adjustment [-]}
\nomenclature[S]{$\bm{h}_b$}{impulse response of BRP $b$}
\nomenclature[S]{$j$}{smart balancing decision index [-]}
\nomenclature[S]{$\alpha,\beta$}{shape parameters of the Beta-distribution}
\nomenclature[S]{$\theta$}{agent parameter}
\nomenclature[S]{$e_\mathrm{RMSE}$}{imbalance energy estimation RMSE [-]}
\nomenclature[S]{$e_\mathrm{half}$}{half width of the robustness interval [-]}
\nomenclature[S]{$e_\mathrm{eff}$}{effective participation [-]}
\nomenclature[S]{$\tau_\mathrm{active}$}{participation fraction [-]}
\nomenclature[S]{$\theta$}{agent parameter}

\nomenclature[E]{E}{exact NRT data}
\nomenclature[E]{I}{interval NRT data}
\nomenclature[E]{F}{future, i.e., unavailable, NRT data}
\nomenclature[E]{rel}{relative}
\nomenclature[E]{ref}{reference}

\printnomenclature
\end{framed}
\end{table*}

The activation of \glspl{frr} is coordinated by the \gls{tso} of the \gls{ca}.
After each \gls{isp}, the \gls{tso} determines the imbalance price based on the amount and incurred costs of the \gls{frr}, and an imbalance settlement is made between the \gls{tso} and each \gls{brp}.
With a single imbalance pricing mechanism, which is the standard mechanism in Europe \citep{entsoeBalGL2018}, \glspl{brp} are penalized for schedule deviations that contributed to the overall \gls{ca} imbalance.
On the other hand, when the imbalance of a \gls{brp} has the opposite sign to the total \gls{ca} imbalance, thus reducing the \gls{ca} imbalance, the corresponding energy is remunerated at the imbalance price.
For a typical \gls{isp} in which the \gls{ca} was short ($E_\mathrm{ACE}<0$), this means that a \gls{brp} $b$ which was short itself ($E_b<0$) would need to pay the \gls{tso}, but would receive money through the imbalance settlement mechanism if it were long ($E_b>0$).
This is reversed for \glspl{isp} during which the \gls{ca} was long ($E_\mathrm{ACE}>0$).

This creates the business case for smart balancing (sometimes called self-balancing, passive control, or passive balancing).
Smart balancing means that a \gls{brp} \emph{intentionally} introduces schedule deviations $E_b$ to reduce the overall \gls{ca} imbalance in order to receive the imbalance remuneration for that schedule deviation. A \gls{brp} can do this in a smart way if it predicts $E_\mathrm{ACE}$ and the \gls{frr} costs with sufficient accuracy and reliability.
In order to estimate these quantities, a \gls{brp} can use private knowledge (e.g., about its own generation and consumption, \gls{frr} signals from the TSO to its own assets, or own PV-/wind-forecasts), or public data.
Some \glspl{tso} (e.g., in the Netherlands and in Belgium) encourage smart balancing by publishing data on the \gls{ca} imbalance and \gls{frr} costs in \gls{nrt} \citep{SummaryBalanceDelta,elia1minute2019}.
This data can help \glspl{brp} to make informed smart balancing decisions and its publication supports fair competition by increasing the transparancy on the current imbalance situation.

\subsection{Adverse Effects of Smart Balancing}\label{sec:intro_adverse}

The analysis in \cref{sec:intro_smart} is based on the \gls{ca} imbalance energy over an \gls{isp}, but the electricity generation and consumption must be balanced not only in terms of energy for each \gls{isp}, but also in terms of power at \textit{each time instance}.
In Europe, \gls{fcr}, \gls{afrr}, and \gls{mfrr} support the system inertia (and the self-regulating effect due to frequency dependency of loads) to achieve this instantaneous {power balance} while limiting the grid frequency deviation $\Delta f$, which reflects the cumulative effect of the power imbalance between the total electrical power in the grid and the total mechanical power \citep{entsoeBalGL2018}.

Because the imbalance settlement mechanism and the incentives for smart balancing are based on an energy balancing approach, \glspl{brp} that want to perform smart balancing are largely indifferent to its effects on frequency stability within an \gls{isp}.
In theory, this can lead to a broad range of adverse effects, caused by the complex interactions between the imbalance pricing mechanism, the \gls{nrt} data, and the unknown control and decision structures used by \glspl{brp} for smart balancing.
For example, smart balancing that is performed at large scale and with fast assets could, in theory, lead to overreactions to the \gls{nrt} data, aggressive smart balancing at the end of an \gls{isp} (when there is less uncertainty about $E_\mathrm{ACE}$ and potential profits), or speculative behaviour with negative impact on frequency stability.

\begin{figure}[t]
    \centering
    \includegraphics[width=0.7\columnwidth]{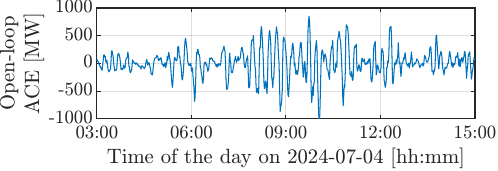}
    \caption{Fast oscillations in the open-loop \gls{ace} in the Netherlands caused by overreactions of \glspl{brp} on \gls{nrt} signals \citep{ENTSOETransparencyPlatform}.}
    \label{fig:NLoscillations}
\end{figure}

Moreover, in practice, smart balancing has been reported to deteriorate frequency stability.
In November 2024, the Dutch \gls{tso} reported persisting \gls{ace} oscillations with magnitudes of up to \qty{1}{GW} due to the very large and fast response of \glspl{brp} to the published \gls{nrt} data, as is illustrated in \cref{fig:NLoscillations}.
The \gls{tso} noted that ``These oscillations~[\ldots] pose a risk to the system stability of the entire continent''~\citep{PowerImbalanceOscillations}.
They reacted by (temporarily) increasing the delay of the \gls{nrt} data from \qty{2}{min} to \qty{5}{min}, increasing the dimensioned \gls{frr} volume, re-designing their \gls{afrr} controller, and limiting the allowed ramp rate of assets \citep{SummaryBalanceDelta}.
These measures seemed effective in reducing the amount and frequency of the oscillations, but did not eliminate them from the Dutch \gls{ca} \citep{TenneTNews20250123}.

\subsection{Motivation, Objective and Innovation}
 
The publication of \gls{nrt} data in the \gls{gcc} region, also known as the German control block, consisting of the four German \glspl{ca}, has been the subject of debate for some time.
The only study that uses a dynamic model to assess the effects of smart balancing was conducted by \citet{RoebenDiss2022}, who focused on the financial impact of different types of \gls{nrt} imbalance data and smart balancing in Germany.
He found that smart balancing would reduce \gls{frr} costs, but could at the same time ``have a negative side effect on the quality of the frequency'', as some frequency-related metrics deteriorated significantly and in some cases smart balancing even led to frequency deviations larger than \qty{200}{mHz} \citep{robenSmartBalancingElectrical2021}.
However, the model used by \citet{RoebenDiss2022} simplifies the behaviour of \glspl{brp} significantly by using the same fuzzy logic controller for the smart balancing decision logic of all BRPs, and by assuming that \glspl{brp} base their smart balancing decision only on the NRT data of the current \gls{isp}.
In reality, it can be expected that each \gls{brp} uses an individual decision logic, and that also NRT data of the previous ISP, as well as private information about the imbalance situation, would be used to come to a smart balancing decision.
The assumptions made by \citet{RoebenDiss2022} can reduce (negative) interactions between BRPs, but can lead to both over- and underestimates of the influence of smart balancing on the frequency stability.

An interesting German report by \citet{hirthSystemstuetzendeBilanzkreisBewirtschaftung} identifies and discusses many possible problems with smart balancing from an economic perspective (relying on energy balances), and comes to an overall favourable evaluation of smart balancing, while recommending further research on the effects of smart balancing within an active ISP. 
Assuming rational BRP decisions, this report dismissed concerns about the type of oscillations later observed in the Netherlands, which highlights the importance of a systematic study on the impact of \gls{nrt} data and smart balancing on the frequency stability of the power system.
Such a systematic study, has -- to the best of the authors' knowledge -- not been carried out yet, neither for the German control block, nor for other power systems.

In this paper, we examine which smart balancing reactions can realistically be expected from \glspl{brp} when \gls{nrt} information is published, and how this affects the system stability, using the \gls{gcc} area as study object.
The focus is exclusively on reactions by assets that are sufficiently fast to perform smart balancing within the current or next \gls{isp} (called physical smart balancing as opposed to financial smart balancing, in which intraday market trading is a part of the smart balancing, by \citet{hirthSystemstuetzendeBilanzkreisBewirtschaftung}).
For the first time in the literature, a dynamic multi-agent system, in which each agent uses a private probabilistic model to estimate the \gls{ca} imbalance and make a smart balancing decision, is created (\cref{sec:mas}).
As environment, the German control block is modelled as a single busbar model \citep{maucherControlAspectsInterzonal2022}, including the system inertia, the load self-regulating effect, as well as the control and activation dynamics of \gls{fcr} and \gls{afrr}.
Scenarios with different types of \glspl{brp} reactions are defined, varying the average \gls{brp} size, estimation accuracy, and risk affinity, and are simulated in a Monte-Carlo setting to evaluate the impact of smart balancing on historical time series of the \gls{gcc} imbalance.
The scenarios that are analysed are presented in \cref{sec:setup}, the results are presented and discussed in \cref{sec:res}, which is followed by the conclusion in\cref{sec:concl}.

\section{Multi-Agent Model} \label{sec:mas}

\subsection{Model Overview} \label{sec:mas_overview}
\begin{figure*}[t]
    \centering
    \includegraphics[width=\textwidth]{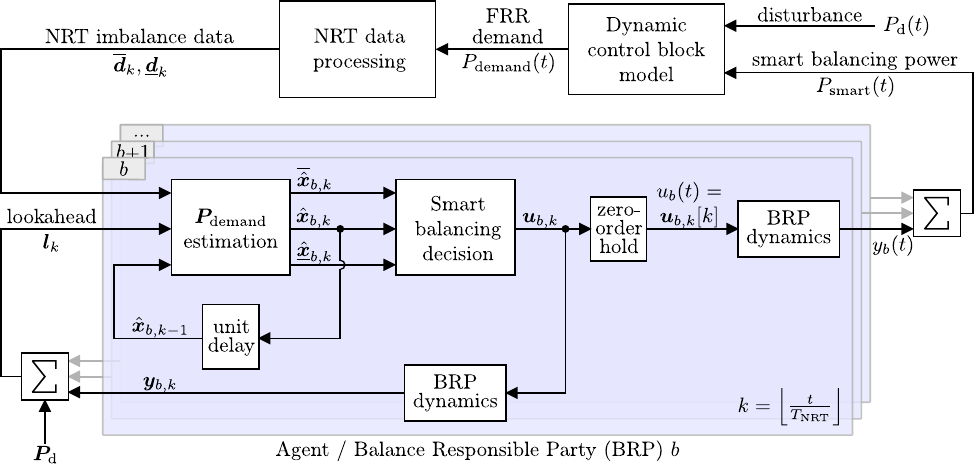}
    \caption{Overview of the multi-agent system.}
    \label{fig:MASoverview}
\end{figure*}

Some parts of the model use discrete time and vector signals, descretized with time step $T_\mathrm{NRT}$, while other parts are in continuous time.
Throughout the paper, continuous time signals are written with their argument in round brackets (e.g., $x(t)$), whereas vector signals are written in bold, and square brackets are used to refer to an element of a vector signal (e.g., $\bm{x}[2]$).
For a simulation with $t\in[0,T_\mathrm{sim}]$, vector signals are column vectors with $N_\mathrm{NRT}$ elements (e.g. $\bm{x} \in \mathbb{R}^{N_\mathrm{NRT}\times1}$), with
\begin{equation}
    N_\mathrm{NRT} = \frac{T_\mathrm{sim}}{T_\mathrm{NRT}}\,.
\end{equation}
All signals that are agent-specific have subscript $b$.
A subscript is also used for vector signals that are updated at discrete times (e.g., $\bm{x}_k$ for a vector created at time $k$, $\bm{x}_{k-1}$ for the vector associated with the previous time step).
Nonitalic sub- and superscripts, as well as tildes, hats and line accents, are descriptive.

\Cref{fig:MASoverview} shows an overview of the multi-agent system.
Each BRP $b$ that can perform smart balancing is modelled as an individual agent, interacting with each other and a common environment, which is the control block in which the BRPs are active.
It is assumed that the agents do not have unintentional schedule deviations, and all unintentional schedule deviations from BRPs that do not perform smart balancing are grouped into a single disturbance signal, $P_\mathrm{d}(t)$.
To make the decision on whether and how to perform smart balancing, each BRP uses a private probabilistic model to estimate the discretized time series of the FRR demand, $\bm{P}_\mathrm{demand}$, and, based on this estimate, chooses a smart balancing reaction to maximize its expected revenues from the imbalance settlement mechanism.
The decision of BRP $b$ depends both on $\hat{\bm{x}}_{b,k}$, which is the estimate of $\bm{P}_\mathrm{demand}$ made at time $k$, and the robustness interval of the estimate, which is defined by an upper and a lower boundary ($\overline{\hat{\bm{x}}}_{b,k}$ and $\underline{\hat{\bm{x}}}_{b,k}$).
The BRP is assumed to only perform smart balancing if positive revenues are expected for the complete robustness interval.
The activation dynamics of the assets of the BRP that physically perform the smart balancing (power plants, storage units,\dots) are also considered in the model.

In order to estimate $\bm{P}_\mathrm{demand}$, the BRPs update their previous estimate  $\hat{\bm{x}}_{b,k-1}$ using a lookahead signal ($\bm{l}_\mathrm{k}$) and the NRT imbalance data ($\overline{\bm{d}}_k$ and $\underline{\bm{d}}_k$), which can be interpreted as feedback from the environment provided by the TSO.
The lookahead signal $\bm{l}_\mathrm{k}$ is the most accurate estimate that is possible at time $k$, considering both the unintentional schedule deviations $\bm{P}_\mathrm{d}$ and the smart balancing actions of other BRPs:
\begin{equation}\label{eq:lookahead}
    \bm{l}_k = \bm{P}_\mathrm{d} + \sum_b \bm{y}_{b,k}\,.
\end{equation}

The combination of these signals makes it possible to vary the degrees to which BRPs rely on the NRT data, trust their private information, and anticipate smart balancing by other BRPs.
Instead of estimating the FRR costs, the modelled agents have access to the marginal price curve according to which the TSO needs to pay for the activation of different amounts of aFRR.
This prevents inaccurate price estimates from negatively influencing the effectiveness of the smart balancing, and makes a robustness interval of the FRR-costs unnecessary, greatly simplifying the model.

\subsection{Environment Model}\label{sec:mas_environment}

\begin{figure}
    \centering
    \includegraphics[width=0.7\columnwidth]{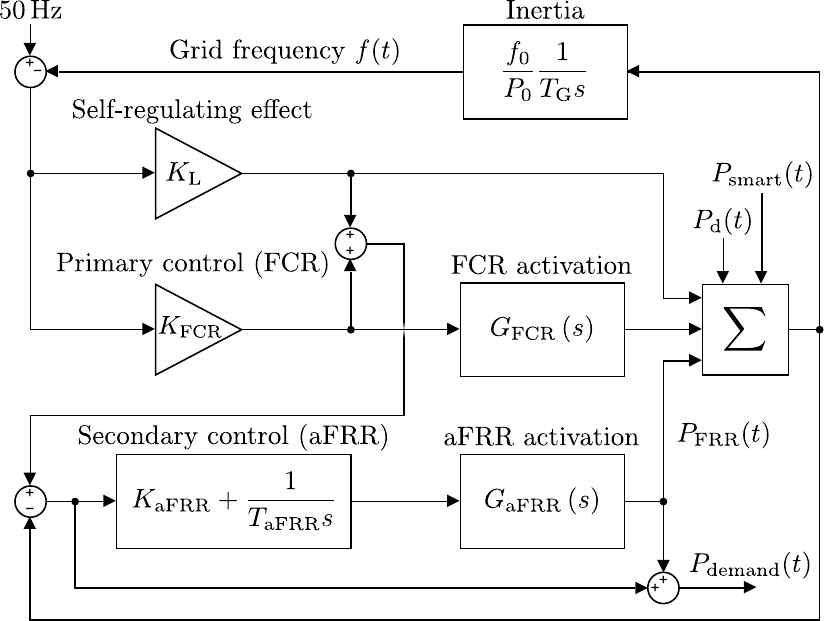}
    \caption{Single busbar model of the control block.}
    \label{fig:MASenvironment}
\end{figure}

Different spatial resolutions and mathematical formulations can be used to model power systems \citep{alayaImpactSpatialResolution2023}.
Single busbar models \citep{maucherControlAspectsInterzonal2022} model \glspl{ca} pointwise. They only consider active power and frequency dynamics, disregarding voltage aspects.
They have a low spatial resolution and are mathematically simple, making them easy to interpret.
This type of model is sufficient for the study of the effects of smart balancing on frequency stability on the \gls{ca} level, and is used to model the control block, effectively considering the four German \glspl{ca} as one single \gls{ca}.
The block diagram of this model is given in \cref{fig:MASenvironment}.
The model contains the control and linearized activation dynamics of the \gls{fcr}, parameterized by $K_\mathrm{FCR}$ and transfer function $G_\mathrm{FCR}(s)$, respectively; the self-regulating effect $K_\mathrm{L}$; the \gls{afrr} control and activation, with parameters $K_\mathrm{aFRR}$, $T_\mathrm{aFRR}$, and transfer function $G_\mathrm{aFRR}(s)$; and the system inertia with time constant $T_\mathrm{G}$.
The parameters of the model are provided in \ref{sec:appEnvironment}.

The control block model takes the unintentional schedule deviations, $P_\mathrm{d}(t)$, the intentional schedule deviations due to smart balancing, $P_\mathrm{smart}(t)$, and the reference frequency $f_0=\qty{50}{Hz}$ as inputs.
Apart from the grid frequency, $f(t)$, the control block model provides the FCR and aFRR power flows, as well as the aFRR demand, $P_\mathrm{demand}(t)$, as outputs.
The implementation of the control block and aFRR control is idealized in \cref{fig:MASenvironment}: because there is no exchange of power with other CAs and the amount of primary control and the self-regulating effect are exactly known, there is no need to use a so-called $K$-factor \citep{entsoeNoteAFRR2018} to obtain the input to the secondary controller.
Instead, the controller input is calculated exactly.
The ACE is not clearly defined within the single busbar model, but $P_\mathrm{demand}(t)$ can be used to define the imbalance situation in the CA, both instantaneously, and over an ISP.
Also $\bm{P}_\mathrm{demand}$, the discretized version of the time series of $P_\mathrm{demand}(t)$, can be used to approximate the FRR energy demand over an ISP:
\begin{align}
    E_\mathrm{demand,ISP}   &= \int_\mathrm{ISP} P_\mathrm{demand}(t')\,\mathrm{d}t'\,, \label{eq:exactEISP} \\
                            &\approx \sum_{k'\in\mathrm{ISP}} \bm{P}_\mathrm{demand}[k']\,.\label{eq:approxEISP}
\end{align}
The ISP is short if $E_\mathrm{demand,ISP}>0$, meaning there is an overall need for positive FRR, and long otherwise.
It is also possible to calculate the marginal costs associated with FRR, for a scalar $P_\mathrm{demand}$.

The second element in the environment in \cref{fig:MASoverview} is the NRT data processing block.
It is assumed that the TSO publishes the average FRR demand for each $T_\mathrm{NRT}=\qty{1}{min}$ period, delayed by $T_\mathrm{delay}$.
The vector signal $\bm{d}_k$ contains the discretized time series of the FRR demand that is available at time $k$,
\begin{equation}\label{eq:timeK}
    k= \left\lfloor\frac{t}{T_\mathrm{NRT}}\right\rfloor\,.
\end{equation}
Elements $\bm{d}_k[i]$ for which imbalance data is available (i.e., $i \leq (t-T_\mathrm{delay})/T_\mathrm{NRT}$) are defined as
\begin{equation}\label{eq:NRTexact}
    \bm{d}_k[i] = \frac{1}{T_\mathrm{NRT}} \int\limits_{(i-1)T_\mathrm{NRT}}^{iT_\mathrm{NRT}} P_\mathrm{demand}(t')\,\mathrm{d}t'\,.
\end{equation}
Elements for which no data is available are undefined.
For the implementation of \cref{eq:NRTexact}, an additional sampling with sampling time $T_\mathrm{TSO}=\qty{4}{s}$ (corresponding with the sampling time used for the calculation of aFRR demand by German TSOs) is used and the integral in \cref{eq:NRTexact} is numerically approximated as the sum over these samples.

It is possible that, instead of publishing $\bm{d}_k$, the TSO pubishes the boundaries of an interval in which $\bm{d}_k$ lies, $[\underline{\bm{d}_k},\overline{\bm{d}_k}]$, called ``traffic light scenarios'' by \citet{RoebenDiss2022}.
This data binning is a generalization of publishing $\bm{d}_k$ exactly, in which undefined elements are replaced by $[-\infty,\infty]$.
Using this generalization, three index ranges of vector elements can be distinguished: elements with exact NRT data (E), interval NRT data (I), or for which no feedback is available yet (future data, F).
This can be formalized as
\begin{subequations}\label{eq:indexRanges}
    \begin{align}
    \mathcal{R}_k^\mathrm{E} &= \Big\{i \,\Big|\, \underline{\bm{d}_k}[i]=\overline{\bm{d}_k}[i]\Big\}\,,\\
    \mathcal{R}_k^\mathrm{I} &= \Big\{i \,\Big|\, \underline{\bm{d}_k}[i]\neq\overline{\bm{d}_k}[i] \neq \infty\Big\}\,,\\
    \mathcal{R}_k^\mathrm{F} &= \Big\{i \,\Big|\, \underline{\bm{d}_k}[i]=-\infty \wedge \overline{\bm{d}_k}[i] = \infty\Big\}\,.
    \end{align}
\end{subequations}

\subsection{Agent Model}\label{sec:mas_agent}
\subsubsection{Imbalance Power Estimation}
The agent model is shown in the blue area in \cref{fig:MASoverview}.
As briefly described in \cref{sec:mas_overview}, a private probabilistic model is used by agent $b$ to estimate $\bm{P}_\mathrm{demand}$, and based on this estimate and the associated robustness interval, the agent decides whether to perform smart balancing at time $k$.
The method does not intend to approximate a single `forecast method' that could be used by a real BRP, but to simulate different types of agent behaviour, allowing variations in the accuracy and consistency of the estimates, as well as in the size (in terms of power available for smart balancing), dynamics and risk-affinity of the agents.
This is achieved by making advantage of the lookahead signal $\bm{l}_k$  as input of the agent model.
Thanks to this signal, which is not available to any BRP in a real world setting, it is possible to simulate estimates of $\bm{P}_\mathrm{demand}$ ranging from very poor to perfect.

The estimate of $\bm{P}_\mathrm{demand}$ at time $k$ relies on this lookahead signal, the NRT feedback signals, $\underline{\bm{d}_k}$ and $\overline{\bm{d}_k}$, and the previous estimate of the agent, $\hat{\bm{x}}_{b,k-1}$.
$\hat{\bm{x}}_{b,k}$ is given by a realisation that is sampled from a truncated normal distribution, which is created from these input signals.
The standard variance of this distribution, $\hat{\bm{\sigma}}_{b,k}$, is approximated to serve as the basis for the robustness interval $[\underline{\hat{\bm{x}}}_{b,k},\overline{\hat{\bm{x}}}_{b,k}]$.

As initial estimate distribution, a multivariate normal distribution with the lookahead signal as mean is used:
\begin{equation}\label{eq:mu_init}
    \bm{\mu}_{b,k} = \bm{l}_k\,.
\end{equation}
Exponential decay in the covariance matrix is used to model the temporal correlation in $\bm{P}_\mathrm{demand}$:
\begin{equation}\label{eq:sigma_init}
    \bm{\Sigma}_{b} = 
    \theta_{\sigma,b}^2 \,\exp\left(-\theta_{\mathrm{d},b} \dfrac{\| i - j \|_1}{N_\mathrm{NRT}}\right)\,.
\end{equation}
The variance and decay parameter, $\theta_{\sigma,b}^2$ and $\theta_{\mathrm{d},b}$, reflect the agents' assumptions on the temporal correlation of $\bm{P}_\mathrm{demand}$.

To obtain the initial estimate of $\bm{P}_\mathrm{demand}$, $\hat{\bm{x}}_{b,-1}$, the normal distribution $\mathcal{N}\left(\bm{\mu}_{b,k}, \bm{\Sigma}_{b}\right)$ is sampled:
\begin{equation}\label{eq:x_init}
   \hat{\bm{x}}_{b,-1} \sim \mathcal{N}(\bm{\mu}_{b,0},\, \bm{\Sigma}_{b})\,.
\end{equation}

When updating the estimate $\hat{\bm{x}}_{b,k}$ for $k\geq0$, $\mathcal{N}\left(\bm{\mu}_{b,k}, \bm{\Sigma}_{b}\right)$ is conditioned on the previous estimate as well as on the NRT data, before sampling the distribution.
It is assumed that the covariance matrix associated with the previous estimate is equal to $\bm{\Sigma}_{b}$, and that $\theta_{\mathrm{w},b}\bm{\Sigma}_{b}$ is the covariance between $\mathcal{N}\left(\bm{\mu}_{b,k}, \bm{\Sigma}_{b}\right)$ and the previous estimate, with $\theta_{\mathrm{w},b}\in[0,1]$ a weighting parameter.
This results in
\begin{equation}\label{eq:mu_weight}
  \bm{\mu}_{\mathrm{w},b,k} = (1 - \theta_{\mathrm{w},b}) \,\bm{\mu}_{b,k} + \theta_{\mathrm{w},b}\, \hat{\bm{x}}_{b,k-1}\,,
\end{equation}
\begin{equation}\label{eq:sigma_weight}
    \bm{\Sigma}_{\mathrm{w},b} = \left(1 - \theta_{\mathrm{w}, b}^{2}\right) \bm{\Sigma}_{b}\,.
\end{equation}

Because the conditioning on the NRT data $\underline{\bm{d}_k}$ and $\overline{\bm{d}_k}$ is different for exact and interval data, $\bm{\mu}_{\mathrm{w},b,k}$ and $\bm{\Sigma}_{\mathrm{w},b}$ are permuted so that elements of $\mathcal{R}_k^\mathrm{E}$, $\mathcal{R}_k^\mathrm{I}$, and $\mathcal{R}_k^\mathrm{F}$ (see \cref{eq:indexRanges}) are grouped:
\begin{equation}\label{eq:mu_weight_EIF}
  \bm{\mu}_{\mathrm{w}} = 
  \begin{pmatrix}
    \bm{\mu}_{\mathrm{w}}^\mathrm{E}\\
    \bm{\mu}_{\mathrm{w}}^\mathrm{I}\\
    \bm{\mu}_{\mathrm{w}}^\mathrm{F}
  \end{pmatrix}\,,
\end{equation}
\begin{equation}\label{eq:sigma_weight_EIF}
  \bm{\Sigma}_\mathrm{w} = 
  \begin{pmatrix}
    \bm{\Sigma}_{\mathrm{w}}^\mathrm{EE} & \bm{\Sigma}_{\mathrm{w}}^\mathrm{EI} & \bm{\Sigma}_{\mathrm{w}}^\mathrm{EF} \\
    \bm{\Sigma}_{\mathrm{w}}^\mathrm{IE} & \bm{\Sigma}_{\mathrm{w}}^\mathrm{II} & \bm{\Sigma}_{\mathrm{w}}^\mathrm{IF} \\
    \bm{\Sigma}_{\mathrm{w}}^\mathrm{FE} & \bm{\Sigma}_{\mathrm{w}}^\mathrm{FI} & \bm{\Sigma}_{\mathrm{w}}^\mathrm{FF}
  \end{pmatrix}\,.
\end{equation}
In these and the following equations, subscripts $b$ and $k$ are implicit.
A similar permutation is applied to the vectors $\underline{\bm{d}}$ and $\overline{\bm{d}}$.

For the exact NRT data, the inequality constraints
\begin{equation}\label{eq:NRTinequality}
    \underline{\bm{d}}\leq\hat{\bm{x}}\leq\overline{\bm{d}}
\end{equation}
simplify to
\begin{equation}\label{eq:estimate_E}
    \hat{\bm{x}}^\mathrm{E} = \overline{\bm{d}}^\mathrm{E}\,,
\end{equation}
which can be considered in the updated imbalance estimate by constructing  a conditional distribution.
This approach cannot be used to account for \cref{eq:NRTinequality} for $\hat{\bm{x}}^\mathrm{I}$.
A truncated multivariate normal distribution ($\mathcal{TN}$), defined by a mean, a covariance matrix, and lower and upper boundaries \citep{daveigaGaussianProcessModeling2012} is therefore used to model $\hat{\bm{x}}^\mathrm{I}$ and $\hat{\bm{x}}^\mathrm{F}$.

The mean and covariance matrix of the truncated normal distribution are obtained by conditioning $\mathcal{N}(\bm{\mu}_\mathrm{w}, \bm{\Sigma}_\mathrm{w})$ on the exact NRT data (\cref{eq:estimate_E}):
\begin{equation}\label{eq:mu_cond}
    \bm{\mu}_\mathrm{c} = 
    \begin{pmatrix}
    \bm{\mu}^\mathrm{I}_\mathrm{c} \\ \bm{\mu}^\mathrm{F}_\mathrm{c}
    \end{pmatrix}
    =
    \begin{pmatrix}
    \bm{\mu}^\mathrm{I}_\mathrm{w} \\ \bm{\mu}^\mathrm{F}_\mathrm{w}
    \end{pmatrix}
    +
    \begin{pmatrix}
        \bm{\Sigma}_{\mathrm{w}}^\mathrm{IE} \\
        \bm{\Sigma}_{\mathrm{w}}^\mathrm{FE} 
    \end{pmatrix}
    {\bm{\Sigma}_{\mathrm{w}}^\mathrm{EE}}^{-1} \left( \overline{\bm{d}}^\mathrm{E} - \bm{\mu}_{\mathrm{w}}^\mathrm{E} \right)\,,
\end{equation}

\begin{subequations}\label{eq:sigma_cond}
\begin{align}
    \bm{\Sigma}_\mathrm{c} &=
    \begin{pmatrix}
        \bm{\Sigma}_\mathrm{c}^\mathrm{II} & \bm{\Sigma}_\mathrm{c}^\mathrm{IF} \\
        \bm{\Sigma}_\mathrm{c}^\mathrm{FI} & \bm{\Sigma}_\mathrm{c}^\mathrm{FF}
    \end{pmatrix}\\
    &=
    \begin{pmatrix}
        \bm{\Sigma}_{\mathrm{w}}^\mathrm{II} & \bm{\Sigma}_{\mathrm{w}}^\mathrm{IF} \\
        \bm{\Sigma}_{\mathrm{w}}^\mathrm{FI} & \bm{\Sigma}_{\mathrm{w}}^\mathrm{FF}
    \end{pmatrix}
    -
    \begin{pmatrix}
        \bm{\Sigma}_{\mathrm{w}}^\mathrm{IE} \\
        \bm{\Sigma}_{\mathrm{w}}^\mathrm{FE}
    \end{pmatrix}
    {\bm{\Sigma}_{\mathrm{w}}^\mathrm{EE}}^{-1}
    \begin{pmatrix}
        \bm{\Sigma}_{\mathrm{w}}^\mathrm{EI} & \bm{\Sigma}_{\mathrm{w}}^\mathrm{EF}
    \end{pmatrix}\,.
\end{align}
\end{subequations}

Using the NRT data as upper and lower limits, the truncated normal distribution can be sampled to obtain $\hat{\bm{x}}^\mathrm{I}$ and $\hat{\bm{x}}^\mathrm{F}$:
\begin{equation}\label{eq:estimate_IF}
    \left( \hat{\bm{x}}^\mathrm{I}, \hat{\bm{x}}^\mathrm{F} \right) 
    \sim
    \mathcal{TN} \left( 
    \bm{\mu}_\mathrm{c}, \,
    {\bm{\Sigma}}_\mathrm{c},\,
    \begin{pmatrix}
        \underline{\bm{d}}^\mathrm{I} & \underline{\bm{d}}^\mathrm{F} 
    \end{pmatrix}
    ,\, 
    \begin{pmatrix}
        \overline{\bm{d}}^\mathrm{I} & \overline{\bm{d}}^\mathrm{F}
    \end{pmatrix}\right)\,.
\end{equation}
If there is no interval feedback, i.e., $\mathcal{R}_k^\mathrm{I}=\emptyset$, \cref{eq:estimate_IF} simplifies to
\begin{equation}\label{eq:estimate_FnoI}
   \hat{\bm{x}}^\mathrm{F} \sim \mathcal{N}(\bm{\mu}_\mathrm{c},\, \bm{\Sigma}_\mathrm{c})\,.
\end{equation}

The upper and lower bounds of the robusness interval are defined as
\begin{subequations}\label{eq:estimate_robustness}
    \begin{align} 
        \overline{\hat{\bm{x}}} &= \hat{\bm{x}} + \theta_\mathrm{z}\,\hat{\bm{\sigma}}\,,\\
        \underline{\hat{\bm{x}}} &= \hat{\bm{x}} - \theta_\mathrm{z}\,\hat{\bm{\sigma}}\,,
    \end{align}
\end{subequations}

with $\hat{\bm{\sigma}}$ an approximation of the standard deviation of the distribution, and $\theta_\mathrm{z}$ an agent-dependent parameter that influences the width of the robustness interval.
For elements with exact NRT feedback, there is no uncertainty about the value of $\hat{\bm{x}}^\mathrm{E}$ (\cref{eq:estimate_E}), hence
\begin{equation}
   \hat{\bm{\sigma}}^{E} = 0\,.
\end{equation}

The exact solution of $\hat{\bm{\sigma}}^\mathrm{I}$ was implemented \citep{lipsTruncatedMultivariate2025} based on \citep{manjunathMomentsCalculationDoubly2021} but was too slow for the Monte-Carlo simulations for which the model is used. 
Instead, an approximation that neglects all cross-correlation (approximating $\bm{\Sigma}_\mathrm{c}$ as diagonal) is used \citep{daveigaGaussianProcessModeling2012}:
\begin{equation}\label{eq:std_I}
  \hat{\bm{\sigma}}^\mathrm{I} = \sqrt{\text{diag}(\bm{\Sigma}_\mathrm{c}^\mathrm{II})}\, \Bigg( 1 +  \frac{\undertilde{\bm{d}}^\mathrm{I} \phi(\undertilde{\bm{d}}^\mathrm{I}) - \widetilde{\bm{d}}^\mathrm{I} \phi(\widetilde{\bm{d}}^\mathrm{I})}{\Phi( \widetilde{\bm{d}}^\mathrm{I} ) - \Phi(\undertilde{\bm{d}}^\mathrm{I})} - \left(\frac{\phi( \widetilde{\bm{d}}^\mathrm{I} ) - \phi(\undertilde{\bm{d}}^\mathrm{I})}{\Phi( \widetilde{\bm{d}}^\mathrm{I} ) - \Phi(\undertilde{\bm{d}}^\mathrm{I})} \right)^{2} \,\Bigg)\,,
\end{equation}
with $\phi$ and $\Phi$ the PDF and CDF of the standard normal distribution, respectively, and $\widetilde{\bm{d}}^\mathrm{I}$ given by
\begin{equation}
    \widetilde{\bm{d}}^\mathrm{I} = \frac
    {\overline{\bm{d}}^\mathrm{I}-\bm{\mu}^\mathrm{I}_\mathrm{c}}
    {\sqrt{\text{diag}(\bm{\Sigma}_\mathrm{c}^\mathrm{II})}}\,,
\end{equation}
and analogous for $\undertilde{\bm{d}}^\mathrm{I}$.

With $\hat{\bm{\sigma}}^\mathrm{I}$ available, an approximation \citep{kotzContinuousMultivariateDistributions2000} of $\hat{\bm{\sigma}}^\mathrm{F}$ is
\begin{equation}
    \hat{\bm{\sigma}}^\mathrm{F} = \sqrt{\text{diag}(\bm{V})}\,,
\end{equation}
using 
\begin{equation}
 \bm{V} = \bm{\Sigma}_\mathrm{c}^\mathrm{FF} - \bm{\Sigma}_\mathrm{c}^\mathrm{FI} \left( {\bm{\Sigma}_\mathrm{c}^\mathrm{II}}^{-1} - {\bm{\Sigma}_\mathrm{c}^\mathrm{II}}^{-1} \text{diag} (\hat{\bm{\sigma}}^\mathrm{I}) \,{\bm{\Sigma}_\mathrm{c}^\mathrm{II}}^{-1} \right) \bm{\Sigma}_\mathrm{c}^\mathrm{IF}\,.
\end{equation}

\subsubsection{Smart Balancing Decision}
It is assumed that the \glspl{brp} either perform no smart balancing or perform smart balancing with the full available capacity.
The smart balancing decision at time $t$ is $u_b(t)=1$ when performing positive smart balancing, $u_b(t)=-1$ when performing negative smart balancing, or $u_b(t)=0$ when not performing smart balancing.
The decision depends on ${\hat{\bm{x}}}_{b,k}$, $\overline{\hat{\bm{x}}}_{b,k}$, and $\underline{\hat{\bm{x}}}_{b,k}$ (see \cref{eq:timeK}) from which estimates for the imbalance price can be derived.
The decision also depends on the costs associated with the activation of smart balancing power.
The imbalance price in the \gls{gcc} is defined based on three price components \citep{NetztransparenzReBAP2023}.
Apart from the \textit{base component}, which depends on the marginal costs associated with the activation of FRR (see \ref{sec:appPrice}), an \textit{incentivising component} is considered, which is coupled with an intraday price index.
The final component is a \textit{scarcity component}, which increases the incentives for BRPs to limit their schedule deviations in ISPs in which the imbalance in the \gls{gcc} is more than \qty{80}{\%} of the dimensioned FRR volume.

$c(\cdot): \mathbb{R}^{N_\mathrm{NRT}\times1} \to \mathbb{R}^{N_\mathrm{NRT}\times1}$ is defined as the function that maps a vector signal of the (estimated) aFRR demand during the simulation to the imbalance price of the respective ISPs, defined by the three price components.
This means that $c(\bm{x})[k']$ is the imbalance price that would apply to the ISP for which $k'\in \mathrm{ISP}$, if the aFRR demand were $\bm{x}$.

A BRP chooses a smart balancing reaction from five possibilities:
\begin{enumerate}
    \item $\bm{\delta}_2$: positive smart balancing until the end of the next ISP (no smart balancing afterwards),
    \item $\bm{\delta}_1$: positive smart balancing until the end of the current ISP,
    \item $\bm{\delta}_0$: no smart balancing,
    \item $\bm{\delta}_{-1}$: negative smart balancing until the end of the current ISP,
    \item $\bm{\delta}_{-2}$: negative smart balancing until the end of the next ISP.
\end{enumerate}
The adjustment $\bm{\delta}_j$ ($j\in\{-2,-1,0,1,2\}$) represents the change applied to the previous smart balancing decision of the BRP, $\bm{u}_{b,k-1}$, such that choosing $\bm{u}_{b,k-1} + \bm{\delta}_j$ results in the effects listed above.

We define $\bm{h}_b$ as the impulse response for the dynamics (see also \cref{sec:mas_agent_activation}) associated with the activation of the smart balancing by BRP $b$, i.e.,
\begin{equation} \label{eq:BRPdynamicsDISCRETE}
    \bm{y}_{b,k} = \bm{h}_b * \bm{u}_{b,k}\,,
\end{equation}
with $\bm{y}_{b,k}$ the activated smart balancing power over time as seen in \cref{fig:MASoverview}.
The revenues obtained by a BRP with impulse response $\bm{h}$, a certain previous decision $\bm{u}$, and decision adjustment $\bm{\delta}$, if the FRR demand were $\bm{x}$, is given as:
\begin{multline} \label{eq:expectedRevenues}
    C\left({\bm{x}},\bm{u},\bm{\delta},\bm{h}\right) = T_\mathrm{NRT}\left(\bm{h} * \left( \bm{u} + \bm{\delta} \right)\right)^\intercal \, c \big( {\bm{x}} - \bm{h} * \bm{u}  \\ + \bm{h} * \left( \bm{u} + \bm{\delta} \right)\big)  - c_\mathrm{smart}\left(\bm{u} + \bm{\delta} \right)\,,
\end{multline}
with $c_\mathrm{smart}\left(\bm{u} + \bm{\delta} \right)$ the costs associated with the activation of smart balancing power for the smart balancing decision $\bm{u} + \bm{\delta}$, which are parametrized by $\theta_{\mathrm{c},b}$ for BRP $b$.
The argument of the imbalance price function $c(\cdot)$ in \cref{eq:expectedRevenues} accounts for the effect that the own reaction has on the FRR demand and the resulting  imbalance price.

Each BRP chooses the adjustment $\bm{\delta}_{j*}$ that results in the highest expected revenues from all reactions for which positive revenues are expected for the complete robustness interval.
The set of reactions for which this holds is
\begin{equation} \label{eq:subsetJ}
    \mathcal{J}_{b,k} = \Big\{j\,\Big|\,j\in\left\{-2,-1,0,1,2\right\} \wedge    C\left({\bm{x}},\bm{u}_{b,k-1},\bm{\delta}_j,\bm{h}_b\right) > 0 \,,\, \forall {\bm{x}} \in \big\{ {\hat{\bm{x}}}_{b,k},\overline{\hat{\bm{x}}}_{b,k},\underline{\hat{\bm{x}}}_{b,k} \big\}\Big\}\,.
\end{equation}
The decision time series of BRP $b$ at time $k$ is then given by
\begin{equation}\label{eq:decisionU}
    \bm{u}_{b,k} = \bm{u}_{b,k-1} + \bm{\delta}_{j*}
\end{equation}
with
\begin{equation}\label{eq:optimalJ}
    j*= 
    \begin{cases} 
        \max\limits_{\mathcal{J}_{b,k}} C\left(\hat{\bm{x}}_{b,k},\bm{u}_{b,k-1},\bm{\delta}_j\right)\,, & \text{if } \mathcal{J}_{b,k} \neq \emptyset\,,\\
        0\,, & \text{if } \mathcal{J}_{b,k} = \emptyset\,.
    \end{cases}
\end{equation}

As seen in \cref{fig:MASoverview}, the decision signal $u_b(t)$ is obtained by applying a zero-order hold to the vector decision signal $\bm{u}_{b,k}$:
\begin{equation}
    u_b(t) = \bm{u}_{b,k}[k]\,.
\end{equation}

\subsubsection{Activation Dynamics}\label{sec:mas_agent_activation}

The activation dynamics of the assets of the BRP that perform the smart balancing are modelled as a first-order system with gain $\theta_{\mathrm{G},b}$ and time constant $\theta_{\mathrm{T},b}$.
The transfer function in Laplace-domain is
\begin{equation}\label{eq:BRPdynamicsCONT}
    \frac{Y_b(s)}{U_b(s)} = \frac{\theta_{\mathrm{G},b}}{\theta_{\mathrm{T},b}\,s+1}\,.
\end{equation}
The equivalent discrete-time dynamics are defined in \cref{eq:BRPdynamicsDISCRETE} using the impulse response $\bm{h}_b$.

\section{Monte-Carlo Simulation Setup} \label{sec:setup}

A Monte-Carlo simulation is used to evaluate the effects of NRT data and smart balancing on the frequency stability of the system.
In each simulation run, different agent and environmental parameters are used, and 100 agents are simulated.
A discrete number of environmental scenarios are defined in \cref{sec:setup_environment}, and different beta distributions from which the agent parameters are sampled are described in \cref{sec:setup_agent}.
With this approach, a wide variety of BRP behaviour can be obtained while simulating a relatively low amount of ISPs (a total of 7680 ISPs were simulated).
Another advantage is that a comparison of statistical properties of ensembles of simulation runs (grouped by the beta distribution used for a certain $\theta$, by historical imbalance time series, $T_\mathrm{delay}$, or by NRT signal type) can be made (see \cref{sec:res_stat}).

The complete method is implemented and simulated using MATLAB/Simulink, and simulations are made for all combinations of scenarios described in \cref{sec:setup_environment} and \cref{sec:setup_agent}.
For the implementation of the sampling from the truncated multivariate normal distribution, the algorithm of \citep{botevNormalLawLinear2017} is used.

\subsection{Environmental Parameters} \label{sec:setup_environment}
The smart balancing behaviour can vary depending on the unintentional schedule deviation $P_\mathrm{d}(t)$,
Four distinct 2-hour periods were selected from the historical \gls{gcc} imbalance time series from 2023 and are used as $P_\mathrm{d}(t)$ for different simulations.
The selected periods and the matching intraday market index that is relevant for the calculation of the imbalance price \citep{NetztransparenzReBAP2023}, are given in \ref{sec:appHistorical}, where they are also briefly discussed.

\begin{figure}
    \centering
    \includegraphics[width=0.7\columnwidth]{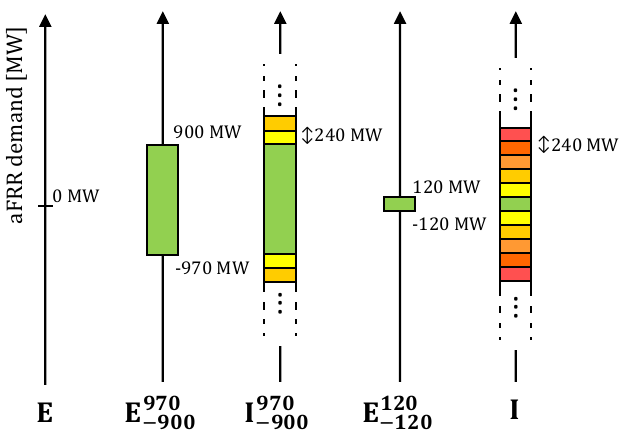}
    \caption{Visualization of the publication methods for the NRT imbalance data.}
    \label{fig:NRTvariations}
\end{figure}

Simulations are performed with ten different NRT signals.
Five different scenarios are defined for the publication of $P_\mathrm{demand}(t)$ and for each of these, scenarios with $T_\mathrm{delay}=\qty{1}{min}$ and $T_\mathrm{delay}=\qty{2}{min}$ are considered.
The NRT data scenarios are visualized in \cref{fig:NRTvariations}.

An abbreviation that encodes the way the signal is published is used to describe the scenarios:
\begin{itemize}
    \item \E: $\bm{d}_k$ (see \cref{eq:NRTexact,eq:indexRanges}) is published exactly.
    \item \Es: $\bm{d}_k$ is published exactly, except for values in the $[-120,120]$ \si{MW} range. For these values, only the interval is published.
    \item \Is: Upper and lower interval boundaries are published for all $\bm{d}_k$, all intervals have a width of \qty{240}{MW}.
    \item \El: $\bm{d}_k$ is published exactly, except for values in the $[-900,970]$ \si{MW} range. For these values, only the interval is published.
    \item \Il: Upper and lower interval boundaries are published as in \Is for values of $\bm{d}_k$ outside of $[-900,970]$ \si{MW}, otherwise the interval $[-900,970]$ \si{MW} is published.
\end{itemize}
The limits of the large interval correspond with approximately \qty{50}{\%} of the average procured aFRR in the \gls{gcc}. 
The limits of the small interval were chosen to represent the expected variance of the \gls{afrr} demand.
Values within this small interval are considered as the random fluctuations of the \gls{afrr} demand.

\subsection{Agent Parameters} \label{sec:setup_agent}
In order to excite different types of dominant smart balancing behaviour in the simulations, the BRP parameters are sampled from predefined beta-distributions in the different runs.
For each BRP parameter, the parameter range and the shape parameters of the beta distribution, $\text{Beta}(\alpha,\beta)$, from which parameters are sampled are given in \cref{tab:agentBetaParameters}.

\begin{table}
    \centering
    \caption{Parameter ranges and beta distributions used to determine the agent parameters for the Monte-Carlo simulation.}
    \label{tab:agentBetaParameters}
    \begin{tabular}{lll}
         \toprule
        Parameter & Range & $(\alpha,\beta)$ \\\midrule
        \multirow{ 2}{*}{$\theta_\mathrm{G}$ (gain)} & \multirow{ 2}{*}{$[10,100]\,$\si{MW}} & $\{(1,10),$ \\[-2pt]
        &&$(10,1)\}$ \\[2pt]
        $\theta_\mathrm{T}$ (time constant) & $\{2,5,10\}\,$\si{min} & uniform \\[2pt]
        \multirow{ 3}{*}{$\theta_\sigma^2$ (variance)} & \multirow{ 3}{*}{$[0.3,1]\cdot \text{var}(P_\mathrm{d})$} & $\{(1,10),$ \\[-2pt]
        &&$(10,10),$ \\[-2pt]
        &&$(10,1)\}$ \\[2pt]
        $\theta_\mathrm{d}$ (decay) & $[0.8,2]$ & $\{(1,1)\}$ \\[2pt]
        $\theta_\mathrm{w}$ (weighting)& $[0.7,0.9]$ & $\{(1,1)\}$ \\[2pt]
        \multirow{ 2}{*}{$\theta_\mathrm{z}$ (std score)} & \multirow{ 2}{*}{$[0.3,3.3]$} & $\{(1,10),$ \\[-2pt]
        &&$(10,1)\}$ \\[2pt]
        $\theta_\mathrm{c}$ (costs) & $\{0\}$ & - \\
        \bottomrule     
    \end{tabular}
\end{table}

The ranges for the gain $\theta_G$ and time constant $\theta_T$ (see \cref{eq:BRPdynamicsCONT}) consider BRPs with different amounts of fast assets that can be used for smart balancing.
The beta distributions from which $\theta_\mathrm{G}$ are sampled are skewed: a simulation either contains mostly BRPs with a small smart balancing gain or BRPs with a very large gain.
In order to make it possible to analyze the results for assets with different dynamics, discrete possibilities for $\theta_T$ are used.
$\theta_T$ is therefore sampled from a discrete uniform distribution.

The parameters $\theta_\sigma^2$ and $\theta_\mathrm{d}$ determine the accuracy and correlation decay of the initial estimate of a BRP (\cref{eq:mu_init,eq:sigma_init,eq:x_init}).
The parameters $\theta_\mathrm{w}$ and $\theta_\mathrm{z}$ affect the weighting between the previous estimate of a BRP and the lookahead signal (\cref{eq:mu_weight,eq:sigma_weight}), and the width of the robustness interval (\cref{eq:estimate_robustness}), respectively.
The ranges for these parameters were determined based on preliminary simulations, but can also be interpreted intuitively.
For example, $\theta_{\mathrm{w}}$ affects the consistency of the estimate.
A BRP $b$ with  $\theta_{\mathrm{w},b} = 1$ has $\hat{\bm{x}}_{b,k} = \hat{\bm{x}}_{b,k-1}$ (see \cref{eq:mu_weight}).
This BRP gives extremely consistent estimates, but does not improve its estimate over time, making $\theta_{\mathrm{w},b} = 1$ unrealistic.
The range in \cref{tab:agentBetaParameters} was found to be realistic; corresponding to very consistent estimates with gradual improvements thanks to the influence of $\bm{l}_b$.

For $\theta_\mathrm{d}$ and $\theta_{\mathrm{w}}$, a continuous uniform distribution is used, whereas simulations are made with the parameters $\theta_\sigma^2$ and $\theta_\mathrm{z}$ sampled from different distributions.
Because multivariate normal distributions are used in \cref{eq:mu_init,eq:sigma_init,eq:x_init}, all BRPs will on average (over many simulation runs) approach the (perfect) lookahead signal as initial estimate.
It is the variance of these estimates that is affected by the  distributions for $\theta_\sigma^2$. 
The three sets of shape parameters listed in \cref{tab:agentBetaParameters} correspond with scenarios in which most BRPs make almost always very accurate, most of the time accurate, or most of the time inaccurate (initial) estimates of $\bm{P}_\mathrm{demand}$.
The two distributions from which $\theta_\mathrm{z}$ is sampled correspond to scenarios in which most BRPs use either a very small or a very large robustness interval.

The costs associated with the smart balancing are set to zero (i.e., $c_\mathrm{smart}(\cdot)=0$) to avoid them from reducing the amount of smart balancing (\cref{eq:expectedRevenues,eq:subsetJ,eq:decisionU,eq:optimalJ}) in the simulations.
This is useful to increase the variation in observed smart balancing behaviour and evaluate the impact on the frequency stability.

\section{Results and Discussion} \label{sec:res}
The results of the Monte-Carlo simulation can be studied for individual runs using time series, on a BRP level, e.g., using scatter plots, or statistically, e.g., using box plots.
Although individual runs are not statistically relevant, each run does present potential smart balancing behaviour in a realistic setting, and any adverse effects on the frequency stability - even when they only occur in a fraction of the runs - could occur in reality under similar conditions.
Simulations with $(\alpha,\beta)=(10,1)$ for $\theta_\mathrm{G}$ and $(\alpha,\beta)=(1,10)$ for $\theta_\mathrm{z}$ systematically resulted in large oscillations in the frequency, caused by very aggressive reactions of BRPs.
Although these effects can occur, these runs are not included in any of the analyses presented here, because they distort the (statistical) representations.

\subsection{Time Series Analysis}\label{sec:res_time}

\begin{figure*}[t]
    \centering
    \includegraphics[width=0.9\textwidth]{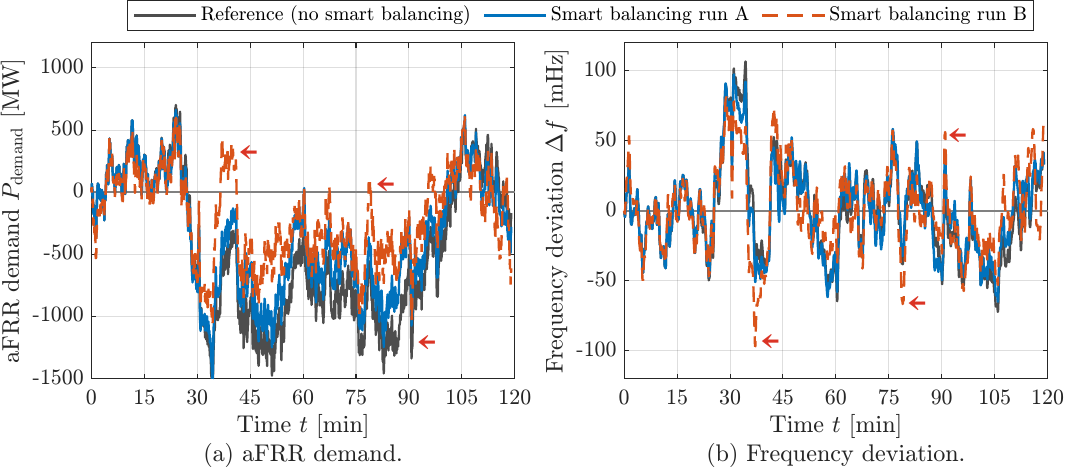}
    \caption{Exemplary time series of two simulation runs with the same $P_\mathrm{d}(t)$, and the reference scenario.}
    \label{fig:res_individualFullRun}
\end{figure*}

\Cref{fig:res_individualFullRun} shows $P_\mathrm{demand}$ and $\Delta f$ for two simulation runs (run A and run B) with the same historical $P_\mathrm{d}(t)$, as well as the results of the reference scenario without smart balancing.
In the simulation run shown with solid blue lines, positive effects from the smart balancing are seen: $P_\mathrm{demand}$ is reduced, especially when $|P_\mathrm{demand}|$ is large in the reference case.
At these times, smart balancing decisions are more likely to be robust, so more BRPs participate.
In this simulation, the smart balancing only has minor impact on the frequency deviation and a positive effect on the \gls{frr} costs and volumes.
Also, for the run shown with dashed orange lines, the smart balancing reduces the overall $P_\mathrm{demand}$.
However, at certain times (marked with red arrows), the frequency deviations increase because of the smart balancing.

\begin{figure}[tb]
    \centering
    \includegraphics[width=0.6\columnwidth]{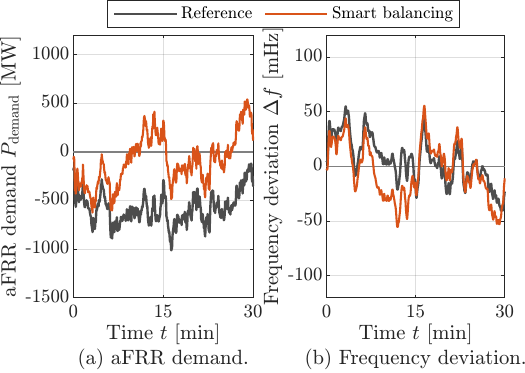}
    \caption{Exemplary time series of two ISPs in a single simulation, with strong smart balancing at the end of the ISPs.}
    \label{fig:res_individualMoreEnd}
\end{figure}

In \cref{fig:res_individualMoreEnd}, the difference between $P_\mathrm{demand}(t)$ in the reference scenario and the shown smart balancing simulation is larger towards the end of the ISPs.
This means that more smart balancing takes place at these times.
Intuitively, it is clear that the NRT data $[\underline{\bm{d}_k},\overline{\bm{d}_k}]$ provides more info about $P_\mathrm{demand}$ at later times.
This implies a smaller uncertainty on $P_\mathrm{demand}$ and the imbalance price towards the end of the ISP, causing more BRPs to participate.
In this simulation, this effect has a negative impact on the frequency margin at the end of both ISPs.

\begin{figure}[tb]
    \centering
    \includegraphics[width=0.6\columnwidth]{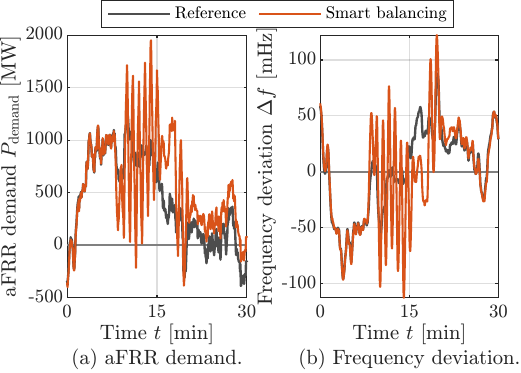}
    \caption{Exemplary time series of two ISPs in a single simulation, showing oscillations due to fast and large smart balancing.}
    \label{fig:res_individualOscillations}
\end{figure}

In \cref{fig:res_individualOscillations}, fast and large oscillations are seen in both $P_\mathrm{demand}$ and $\Delta f$. 
These oscillations are caused by the quasi-simultaneous decision of many BRPS to perform smart balancing, resulting in a strong overreaction (similar to the overreactions marked with red arrows in \cref{fig:res_individualFullRun}, but with a larger amplitude).
Although the individual effect of the smart balancing by each BRP would be good for the system, the combined effect negatively affects \gls{gcc} imbalance and the frequency stability.
Thanks to the NRT data, the BRPs become aware of this overreaction, and overcompensate again by performing smart balancing in the opposite direction, resulting in oscillations which they fail to effectively control until the environmental context changes.

\subsection{BRP-level Analysis}\label{sec:res_BRP}

\begin{figure}[t]
    \centering 
    \includegraphics[width=0.6\columnwidth]{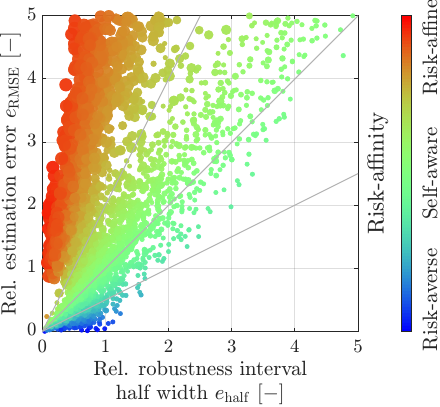}
    \caption{Evaluation of the smart balancing performance of BRPs for simulations with NRT data of type \E and $T_\mathrm{delay}=\qty{60}{s}$.}\label{fig:res_BRPestimateE}
\end{figure}

\begin{figure}
\centering
\includegraphics[width=0.7\columnwidth]{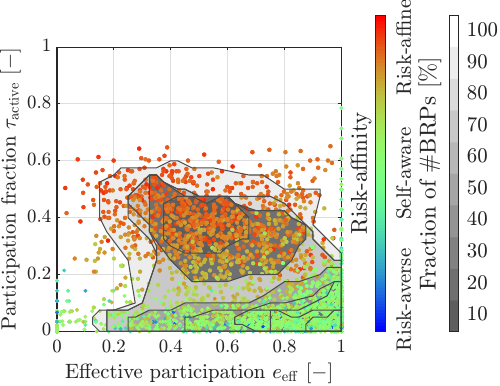}
    \caption{Evaluation of the smart balancing performance of BRPs for simulations with NRT data of type \E and $T_\mathrm{delay}=\qty{60}{s}$.}
    \label{fig:res_BRPsmartE}
\end{figure}

Two figures that show a multidimensional analysis of the behaviour of the BRPs are presented in \cref{fig:res_BRPestimateE,fig:res_BRPsmartE}.
In the figures, each BRPs of simulations with exact (\E) NRT data and $T_\mathrm{delay}=\qty{60}{s}$ are plotted as individual points.
The figures are used to evaluate the performance of the BRPs and how this relates to the effectiveness of the smart balancing performance.

In order to evaluate the estimation performance, \cref{eq:approxEISP} is evaluated during the simulation to approximate $E_{\mathrm{demand,ISP},k}$ of the current ISP.
Analogously, $\hat{E}_{\mathrm{demand,ISP},b,k}$, $\overline{\hat{E}}_{\mathrm{demand,ISP},b,k}$, and $\underline{\hat{E}}_{\mathrm{demand,ISP},b,k}$ are calculated based on $\hat{\bm{x}}_{b,k}$, $\overline{\hat{\bm{x}}}_{b,k}$, and $\underline{\hat{\bm{x}}}_{b,k}$, respectively.
From these quantities, we define the relative root-mean-square error (RMSE),
\begin{equation}\label{eq:ermse}
    e_{\text{\scriptsize RMSE},b} = \frac{ \sqrt{\sum_k\left(\hat{E}_{\mathrm{demand,ISP},b,k} - E_{\mathrm{demand,ISP},k} \right)^2}}{E_{\mathrm{demand,ISP},k}}\,,
\end{equation}
and the average relative half width of the robustness interval,
\begin{equation}\label{eq:ehalf}
    e_{\text{\scriptsize half},b} = \sum_k \frac{1}{2}\frac{\overline{\hat{E}}_{\mathrm{demand,ISP},b,k}-\underline{\hat{E}}_{\mathrm{demand,ISP},b,k}}{E_{\mathrm{demand,ISP},k}}\,.
\end{equation}

$e_{\text{\scriptsize RMSE},b}$ is an evaluation metric for the average goodness of the estimation of BRP $b$, whereas $e_{\text{\scriptsize half},b}$ quantifies the width of the average robustness interval.
Their relation $e_{\text{\scriptsize RMSE},b}/e_{\text{\scriptsize half},b}$ is an indicator for the risk-affinity of the BRP: if  $e_{\text{\scriptsize RMSE},b} \approx e_{\text{\scriptsize half},b}$, the robustness interval of BRP $b$ correctly accounts for the inaccuracy of its FRR demand estimate.
Such a BRP is called \textit{self-aware}.
However, if $e_{\text{\scriptsize RMSE},b}>e_{\text{\scriptsize half},b}$, the robustness interval is small compared with the error made in the estimate.
This corresponds with \textit{risk-affine} BRPs, that will perform smart balancing even with a high uncertainty on the outcome of the action.
The opposite corresponds to \textit{risk-averse} BRPs.

In \cref{fig:res_BRPestimateE}, a scatter plot of these quantities is shown.
The colours in the figure indicate the risk-affinity of the BRPs, ranging from blue (risk-averse) over green (self-aware) to red (risk-affine). 
Some BRPs (\qty{2.9}{\%} in this case) are not displayed because their RMSE was too large to fit on the figure.
The size of the points corresponds to the participation fraction $\tau_{\mathrm{active},b}$, which is the fraction of the time for which BRP $b$ decided to perform smart balancing (i.e., the time for which $j* \neq 0$, see \cref{eq:optimalJ}).
The risk-affine BRPs are typically larger in the scatter plot, meaning that they perform smart balancing more often than self-aware and risk-averse BRPs, which matches the expectations.

In \cref{fig:res_BRPsmartE}, a scatter plot of the same BRPs with the same colors is shown.
On the y-axis, the participation fraction $\tau_{\mathrm{active},b}$ is given, meaning that BRPs at the bottom of the plot only perform smart balancing for an infinitesimal amount of time, whereas BRPs at the top of the plot continuously perform smart balancing.
The x-axis shows $e_{\mathrm{eff},b}$, which is the fraction of smart balancing energy that effectively reduces the system imbalance (i.e., supports the system), evaluated per ISP.
BRPs with $e_{\mathrm{eff},b}>0.5$ are overall supporting the system and reduce the total FRR volume in the simulation, whereas BRPs with $e_{\mathrm{eff},b}<0.5$ worsen the overall system imbalance.

It is seen that the risk-affine BRPs form a cluster in the middle of \cref{fig:res_BRPsmartE}: these BRPs perform smart balancing for about half of the simulation and have varying success in reducing the system imbalance, as they have a large variation in $e_{\mathrm{eff},b}$.
The self-aware (and risk-averse) BRPs cluster in the bottom right of the plot: they only perform smart balancing for a small fraction of the time, and reduce the system imbalance while doing so.
These BRPs ``wait'' until the available NRT data results in $\bm{P}_\mathrm{demand}$ estimates with sufficiently low uncertainty before participating, which is typically only the case towards the end of an ISP.
This type of behaviour is associated with the effect seen in \cref{fig:res_individualMoreEnd}.
Risk-affine BRPs do not wait for this additional information, and participate earlier in the ISP, accepting the higher risks associated with this strategy.
They deliver more smart balancing energy than the risk-averse and self-aware BRPs, but also deteriorate the system imbalance more often.

\begin{figure}[t]
    \centering 
    \includegraphics[width=0.6\columnwidth]{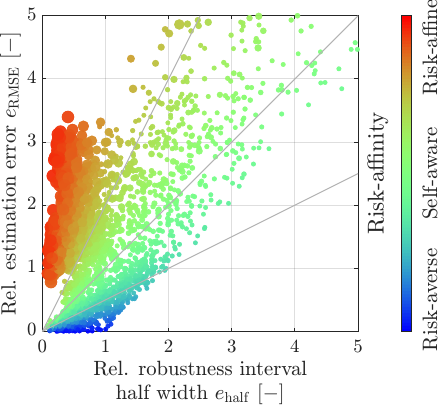}
    \caption{Evaluation of the smart balancing performance of BRPs for NRT data of type \El and $T_\mathrm{delay}=\qty{60}{s}$.}\label{fig:res_BRPestimateEl}
\end{figure}

\begin{figure}
\centering
\includegraphics[width=0.7\columnwidth]{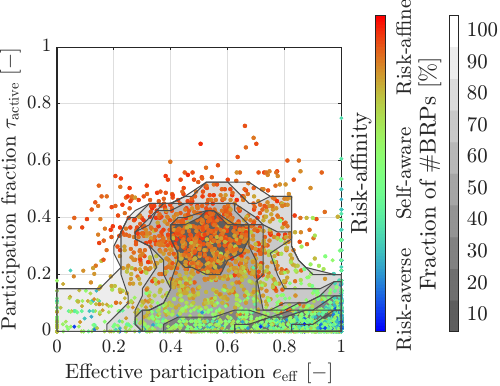}
    \caption{Evaluation of the smart balancing performance of BRPs for NRT data of type \El and $T_\mathrm{delay}=\qty{60}{s}$.}
    \label{fig:res_BRPsmartEl}
\end{figure}

The estimation performance and smart balancing performance of the BRPs vary for different types of NRT feedback and $T_\mathrm{delay}$.
The biggest difference is seen between exact NRT data and NRT data that is published in large intervals (\El and \Il).
\Cref{fig:res_BRPestimateEl,fig:res_BRPsmartEl} show the results for NRT data of type \El and $T_\mathrm{delay}=\qty{60}{s}$.
In \cref{fig:res_BRPestimateEl}, it is seen that there are less risk-affine BRPs with large estimation errors ($e_{\text{\scriptsize RMSE}}$) compared to \cref{fig:res_BRPestimateE}.
This indicates that with exact NRT data, BRPs are more likely to be overconfident about the goodness of their estimates within the modelling framework, resulting in robustness intervals that do not sufficiently account for, e.g., an increased variability of the aFRR demand in the near future.

Although this results in lower participation factors for the BRPs, as is seen in \cref{fig:res_BRPsmartEl}, this does not improve the effectiveness of the smart balancing.
Instead, the cluster of risk-affine BRPs in \cref{fig:res_BRPsmartEl} is concentrated around $e_{\mathrm{eff}}=0.5$ for \El, compared to the results for \E where the cluster was wider, with its centre around $e_{\mathrm{eff}}=0.55$ to $0.6$.
This means that, on average, the combined effect of smart balancing by risk-affine BRPs has a positive effect on the system imbalance when NRT data is published exact, but does not have this effect if large intervals are used for the communication of NRT imbalance data.
It also means that less smart balancing is performed in total when data is published in large intervals.
The results for \Es and \Is are similar to those shown for \E, whereas the results for \Il are similar to the results for \El.

Using the same type of analysis, it was also found that the communication of NRT data in small intervals (\Es and \Is) has the additional benefit that the average active time of the self-aware cluster is larger and associated with a higher success rate, increasing the (overall small) contribution of self-aware BRPs to reducing the system imbalance. 

\begin{figure}
\centering
\includegraphics[width=0.7\columnwidth]{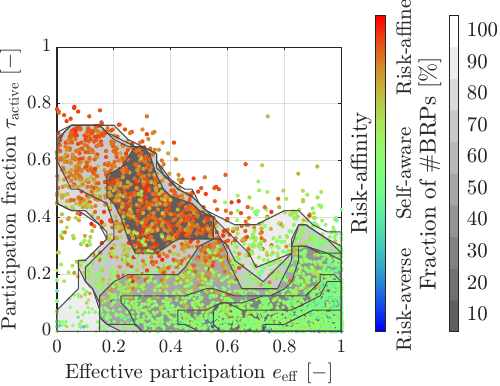}
    \caption{Evaluation of the smart balancing performance of BRPs for simulations with NRT data of type \E and $T_\mathrm{delay}=\qty{60}{s}$, for simulations with $\bm{l}_k = \bm{P}_\mathrm{d}$.}
    \label{fig:res_BRPsmartE_nocompetition}
\end{figure}

Finally, it is possible to analyse the impact of the lookahead signal on the performance of the BRPs. 
An additional Monte-Carlo simulation was made for \E NRT data with $T_\mathrm{delay}=\qty{60}{s}$, in which the BRPs did not have access to the planned smart balancing reactions by other BRPs, meaning $\bm{l}_k = \bm{P}_\mathrm{d}$ instead of \cref{eq:lookahead}.
The smart balancing performance for this Monte-Carlo simulation is shown in \cref{fig:res_BRPsmartE_nocompetition}, where it is seen that both self-aware and risk-affine BRPs perform worse and perform smart balancing more often in this case, as they see a lot of opportunity for smart balancing, but forget to factor in the effect of competitors reacting similarly.
Scenarios such as this could occur in reality when a TSO starts to publish NRT data, so that BRPs do not yet know how the competition performs smart balancing.

\subsection{Statistical Analysis}\label{sec:res_stat}

The differences in effectiveness of the smart balancing discussed in \cref{sec:res_BRP} also have consequences for the frequency stability. 
Different key performance indicators regarding FRR amounts and costs, as well as frequency stability are therefore statistically evaluated for the ensembles of simulation runs for each of the NRT data scenarios (using the default lookahead signal as per \cref{eq:lookahead}).
The results are presented as box-and-whisker plots, characterized by the median (red line), the interquartile range or IQR (a blue box containing \qty{50}{\%} of the simulation results), the whiskers (dashed black lines, containing the tails of the distribution using the IQR-method) and outliers (red crosses). 
Outliers should be interpreted as extreme values (and not as faulty or irrelevant data points).

\subsubsection{Effects on the Balancing Energy}

\begin{figure}[tb]
    \centering
    \includegraphics[width=0.6\columnwidth]{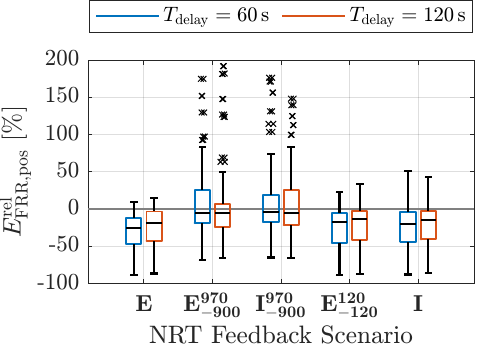}
    \caption{Relative change of the amount of activated positive balancing energy.}
    \label{fig:res_EaFRRpos}
\end{figure}

\begin{figure}[tb]
    \centering
    \includegraphics[width=0.6\columnwidth]{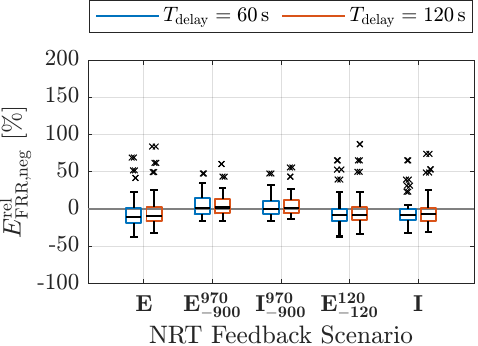}
    \caption{Relative change of the amount of activated negative balancing energy.}
    \label{fig:res_EaFRRneg}
\end{figure}

For each run, the amount of activate positive balancing energy, $E_\mathrm{FRR,pos}$ is calculated as the integral of the positive share of $P_\mathrm{FRR}$, seen in \cref{fig:MASenvironment}.
Analogously, $E_\mathrm{FRR,neg}$ is calculated.
In order to interpret these quantities, they are compared to the value obtained for a \textit{reference} (ref) run with the same environmental parameters but no smart balancing, resulting in the relative change in activated FRR energy 
\begin{equation}\label{eq:relativeResults}
    E_\mathrm{FRR,pos}^\mathrm{rel} = \frac{E_\mathrm{FRR,pos}-E_\mathrm{FRR,pos}^\mathrm{ref}}{E_\mathrm{FRR,pos}^\mathrm{ref}}\,,
\end{equation}
and analogous for $E_\mathrm{FRR,neg}^\mathrm{rel}$.

$E_\mathrm{FRR,pos}^\mathrm{rel}$ and $E_\mathrm{FRR,neg}^\mathrm{rel}$ are statistically evaluated for the different scenarios of NRT imbalance data and presented in \cref{fig:res_EaFRRpos,fig:res_EaFRRneg}.
The outliers in these plots are caused by small reference values in one of the historical datasets used as a basis for the simulation.
The results show no large differences between the 60s and 120s delay scenarios. 
Also, for positive and negative balancing energy, the results are similar. 
However, between the different NRT feedback scenarios, significant differences are seen. 
For the positive aFRR, smart balancing (on average) results in a reduction of the activated aFRR energy for all scenarios. 
If the NRT feedback is published exactly, the energy reduction is the largest. 
If the NRT information is published with a large central interval (\El, \Il), the median of the box plots lies around zero, meaning that only in half of the simulations in this ensemble, the positive aFRR energy was reduced. 
A possible explanation is that, with less accurate information, the BRPs rely more on their own, possibly inaccurate, estimation.
Also, many BRPs might react simultaneously at a moment when the feedback signal leaves the large central interval, potentially causing a common overreaction.
The results are similar for the amount of negative balancing energy.

\begin{figure}[tb]
    \centering
    \includegraphics[width=0.6\columnwidth]{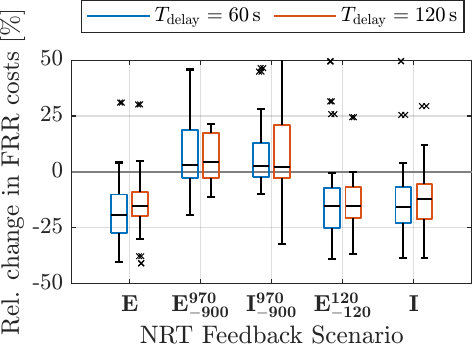}
    \caption{Relative change of the aFRR costs.}
    \label{fig:res_aFRRcosts}
\end{figure}

The relative change of the costs associated with the activation of aFRR are shown in \cref{fig:res_aFRRcosts}.
When the NRT data is published as \E, \Es, or \Is, the costs for aFRR lower on average.
For \El and \Il, the increased amounts of aFRR seen in \cref{fig:res_EaFRRpos,fig:res_EaFRRneg} lead to a significant increase of the total costs, corresponding with positive values in \cref{fig:res_aFRRcosts}.
These results add an important nuance to the results presented by
\citet{robenSmartBalancingElectrical2021}, who found that smart balancing always lowers activated FRR amounts and costs for various types of NRT imbalance data, and that exact NRT data (\E) lowers the costs the most.
Our study shows that the impact on FRR amounts and costs also strongly depends on the decision logic, and that smart balancing can increase the amount of activated FRR energy and associated costs in some cases.
These results are not contradictory, as \citet{robenSmartBalancingElectrical2021} did not consider variations in the decision logic of the BRPs.

\subsubsection{Effects on the Frequency Stability}
The effects of smart balancing on the frequency stability can be studied by analysing the mean frequency deviation $\Delta f_\mathrm{mean}$, the maximum and minimum frequency deviation, $\Delta f_\mathrm{max}$ and $\Delta f_\mathrm{min}$, and the variability of the frequency deviation, for which the standard deviation over different time intervals is calculated: $\Delta f_\mathrm{std,15min}$ and $\Delta f_\mathrm{std,1min}$.

\begin{figure}[tb]
    \centering
    \includegraphics[width=0.6\columnwidth]{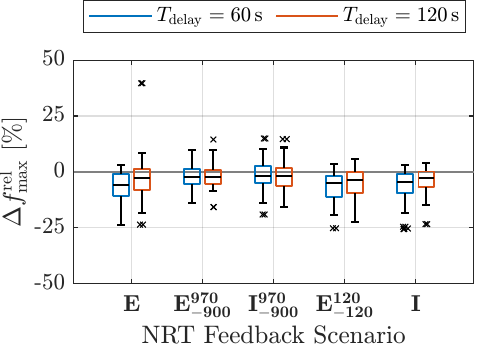}
    \caption{Relative change of the maximum frequency deviation.}
    \label{fig:res_dfmax}
\end{figure}

\begin{figure}[tb]
    \centering
    \includegraphics[width=0.6\columnwidth]{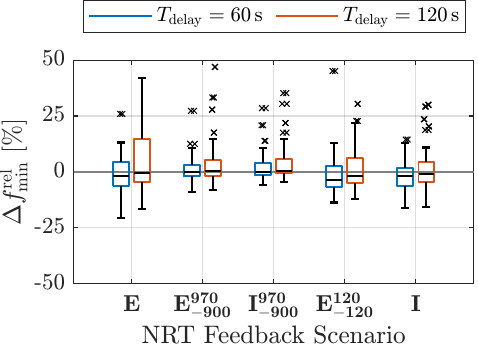}
    \caption{Relative change of the minimum frequency deviation.}
    \label{fig:res_dfmin}
\end{figure}

\begin{figure}[tb]
    \centering
    \includegraphics[width=0.6\columnwidth]{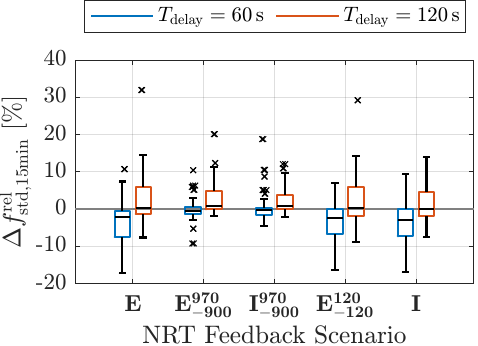}
    \caption{Relative change of the 15-minute standard deviation of the frequency deviation.}
    \label{fig:res_dfstd15}
\end{figure}

\begin{figure}[tb]
    \centering
    \includegraphics[width=0.6\columnwidth]{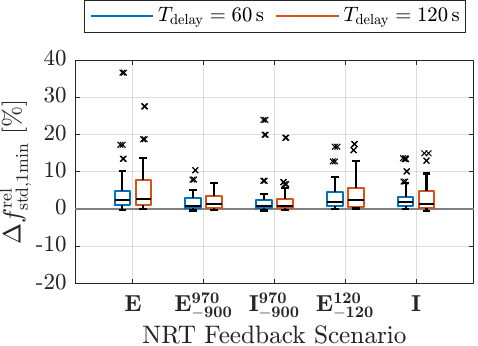}
    \caption{Relative change of the 1-minute standard deviation of the frequency deviation.}
    \label{fig:res_dfstd1}
\end{figure}

Because the frequency is a controlled quantity and the secondary controller eliminates stationary deviations, smart balancing does not significantly influence $\Delta f_\mathrm{mean}$.
The other quantities are shown in \cref{fig:res_dfmax,fig:res_dfmin,fig:res_dfstd15,fig:res_dfstd1}, relative to a reference scenario, analogous to \cref{eq:relativeResults}.
For all these graphs, improvements to the frequency stability are made for values smaller than zero, and the frequency stability deteriorates for values larger than zero.

It is seen that $\Delta f_\mathrm{max}$ typically decreases compared with the reference simulation thanks to the smart balancing, which is good for the system. 
However, $\Delta f_\mathrm{min}^\mathrm{rel}$ has median values close to \qty{0}{\%}, and the upper limit of the IQR is significantly larger than zero when certain types of NRT imbalance data are published.
When this is the case, the frequency margin for negative frequency deviations is significantly worsened because of the smart balancing.
The highest value for the upper IQR limit is seen when exact NRT data is published with $T_\mathrm{delay}=\qty{120}{s}$.
The NRT feedback scenarios with a large central interval (\El, \Il) typically have the smallest IQRs, which are also the closest around zero, indicating that smart balancing has a smaller overall effect on the frequency margin for these scenarios.
This agrees with the findings from \cref{sec:res_BRP} that less smart balancing occurs for \El and \Il types of NRT data, and that the smart balancing by risk-affine BRPs does not reduce the system imbalance, as it is concentrated around $e_{\text{\scriptsize half}} = 0.5$.
The difference between the results for positive and negative frequency deviations (\cref{fig:res_dfmax,fig:res_dfmin}) could be linked with the asymmetry in the aFRR marginal prices for positive and negative aFRR, which determine the incentives for smart balancing through the imbalance price (see \ref{sec:appPrice}).

The frequency variability, shown in \cref{fig:res_dfstd15,fig:res_dfstd1}, is another indicator for the system stability, as sharp variations cause stronger control actions.
It is seen that smart balancing systematically worsens the short-term variability, $\Delta f_\mathrm{std,1min}$, relative to the reference scenario without smart balancing.
The variability evaluated per ISP, $\Delta f_\mathrm{std,15min}$, generally improves if $T_\mathrm{delay}=\qty{60}{s}$, but it becomes worse if smart balancing is performed with $T_\mathrm{delay}=\qty{120}{s}$.
In control systems, large dead times are known to worsen control performance and potentially cause oscillations. 
When interpreted in this context, an even longer delay (potentially also a delay in the activation of the smart balancing energy) could further worsen the frequency variability. 
Interestingly, for scenarios with $T_\mathrm{delay}=\qty{60}{s}$, positive effects for $\Delta f_\mathrm{std,15min}$ are coupled with negative effects on $\Delta f_\mathrm{std,1min}$.

\section{Conclusions and Outlook}\label{sec:concl}

This study is the first to systematically examine which smart balancing reactions can realistically be expected when \gls{nrt} imbalance data is published and how this affects the frequency stability.
A Monte-Carlo simulation using a dynamic multi-agent model was performed to analyse the effects of smart balancing on the frequency stability in the \gls{gcc} area.
All major power balancing mechanisms, \gls{nrt} feedback about the imbalance situation, and structures to enable smart balancing by \glspl{brp} are included in the model.
The model is simulated for multiple thousand \glspl{isp} with different agent and environmental parameters, after which the simulations are evaluated using statistical analysis as well as analysis of individual time series and BRPs.

It is found that smart balancing using \gls{nrt} imbalance data information is beneficial for the system in many scenarios, but worsens the frequency stability under certain circumstances that could occur in reality.
When \gls{nrt} imbalance data is published, \glspl{brp} typically perform smart balancing only for a small fraction of the total time.
This indicates that \glspl{brp} actively use the \gls{nrt} data to re-evaluate their decision.
Smart balancing typically reduces the amount of activated \gls{frr} energy and associated costs.
However, it is also possible that smart balancing leads to increased FRR energy (and costs), especially when the \gls{nrt} data is published with coarse granularity (\El and \Il).
A more exact publication of the \gls{nrt} data is beneficial to reduce these adverse effects.

It is found that, under realistic assumptions, smart balancing could negatively impact the frequency stability in different ways.
Smart balancing typically increases the (short-term) frequency variability, examples are seen in \cref{fig:res_individualMoreEnd,fig:res_individualOscillations}, which can increase the required frequency control effort and FCR demand.
The positive frequency margin typically improves thanks to smart balancing.
However, the negative frequency margin can decrease (an example is shown in \cref{fig:res_individualFullRun}) significantly, most notably when there is a larger delay in the \gls{nrt} data.

In general, less adverse effects are seen when the \gls{nrt} imbalance data is published with a fine granularity (either exact or using small interval ranges) and with a small delay.
A possible explanation is that, when the NRT feedback is provided as large intervals or with a high time delay, a change or worsening of the imbalance only becomes evident after some time or after a large error has accumulated.
As a result, \glspl{brp} might perform smart balancing based on less accurate data, or many \glspl{brp} might react simultaneously, causing overreactions.
These conclusions are in line with the findings of \citet{robenSmartBalancingElectrical2021}, who did not study the impact of time delay in the NRT signal, but also found that publishing the imbalance data using large intervals has the worst effect on the frequency stability for the specific smart balancing logic they studied.

In future work, the decision logic of the agents could be adapted to also consider less benevolent \gls{brp} behaviour.
It is expected that the \glspl{brp} could try to actively increase the imbalance price in order to increase the profits obtained from smart balancing.
This could be studied by using a unit commitment optimization for each individual agent, which would significantly increase the computational complexity of the model.
Further research could also investigate the impact of different imbalance pricing mechanisms on the smart balancing and the frequency stability using the presented model, as a stochastic extension of the research presented by \citet{robenSmartBalancingElectrical2021}.

\section{CReDiT Author Statement}
\textbf{Johannes Lips}: Methodology, Software, Visualization, Writing -- Original Draft
\textbf{Boyana Georgieva}: Software, Visualization, Writing -- Review \& Editing
\textbf{Dominik Schlipf}: Conceptualization, Data Curation, Writing -- Review \& Editing
\textbf{Hendrik Lens}: Supervision, Writing -- Review \& Editing

\section{Acknowledgements}
The authors acknowledge the financial support provided by Transnet BW, which made this research possible.

\FloatBarrier
\bibliographystyle{unsrtnat}   
\bibliography{zotero_library.bib}

\appendix
\setcounter{table}{0}
\setcounter{figure}{0}
\section{Model Details}\label{sec:app}
\subsection{Environmental Parameters}\label{sec:appEnvironment}
\begin{minipage}{\columnwidth}
\centering{
    \captionof{table}{Parameters of the single busbar model.}
    \label{tab:modelParameters}
    \begin{tabular}{ll}
    \toprule
    \textbf{Parameter} & \textbf{Value} \\
    \midrule
    \multicolumn{2}{l}{\textbf{Inertia}} \\
    Reference frequency & $f_{0} = 50 \, \text{Hz}$ \\
    Reference system load & $P_{0} = 300 \, \text{GW}$ \\
    Grid inertia & $T_{\mathrm{G}} = 12 \, \text{s}$ \\  
    \midrule
    \multicolumn{2}{l}{\textbf{Primary Control: FCR}} \\
    \multicolumn{2}{l}{Limited symmetrically at $1\,\text{GW}$} \\
    \midrule
    \multicolumn{2}{l}{\textbf{Secondary Control: aFRR}} \\
    Gain & $K_{\mathrm{aFRR}} = 0.1$ \\
    Integrator time constant & $T_{\mathrm{aFRR}} = 250\, \text{s}$ \\
    \midrule
    \multicolumn{2}{l}{\textbf{Self-regulating effect:}} \\
    \multicolumn{2}{l}{\textbf{frequency dependency of loads}} \\
    Gain & $K_{\mathrm{L}} = 1.75\%/\%$ \\
    Reference control block load &  $80 \, \text{GW}$ \\
    \bottomrule
\end{tabular}

    }
\end{minipage}
\begin{minipage}{\columnwidth}
\centering{
    \captionof{table}{Transfer functions of the single busbar model.}
    \label{tab:modelTFs}
    \begin{tabular}{ll}
    \toprule
    \textbf{Parameter} & \textbf{Value} \\
    \midrule
      FCR activation dynamics & $G_{\mathrm{FCR}}(s) = \frac{9s+1}{(7.5s+1)^2}$\\
     aFRR activation dynamics & $G_{\mathrm{aFRR}}(s) = \frac{1}{20s+1}$\\
    \bottomrule
\end{tabular}
    }
\end{minipage}

\subsection{Marginal aFRR Price Curve}\label{sec:appPrice}
\begin{minipage}{\columnwidth}
\centering{
    \includegraphics[width=0.6\columnwidth]{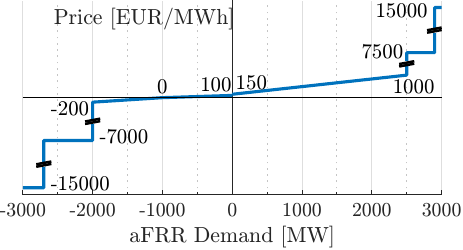}
    \captionof{figure}{Marginal aFRR price curve used in the simulations.}
    \label{fig:marginalaFRR}
    }
\end{minipage}

\subsection{Historical Imbalance Data}\label{sec:appHistorical}
    
\begin{figure}[H]
    \centering
    \subfloat[\centering {Small amplitude variations.}\label{fig:historic1}]{\includegraphics[width=0.7\columnwidth]{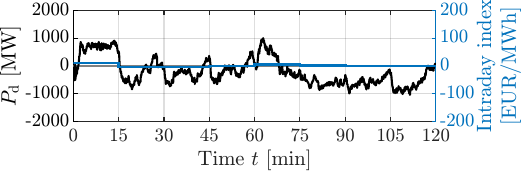}}\\
    \subfloat[\centering {Slowly varying variations with sudden sign reversal.}\label{fig:historic147}]{\includegraphics[width=0.7\columnwidth]{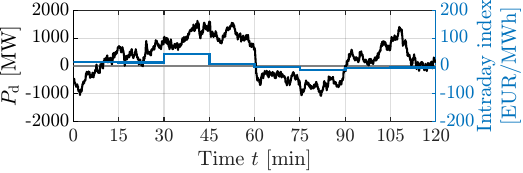}}\\
    \subfloat[\centering {Quickly varying large amplitude variations.}\label{fig:historic156}]{\includegraphics[width=0.7\columnwidth]{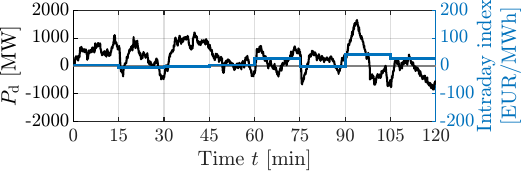}}\\
    \subfloat[\centering {Slowly varying large amplitude variations.}\label{fig:historic175}]{\includegraphics[width=0.7\columnwidth]{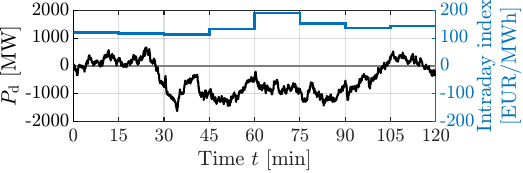}}
    \caption{Historical FRR demand for the \gls{gcc} for selected periods in 2023.}
    \label{fig:historicDemand}
\end{figure}

\end{document}